\documentstyle[preprint,prb,aps]{revtex}
\begin{document}
\draft
\title{First order transition from antiferromagnetism to
ferromagnetism in Ce(Fe$_{0.96}$Al$_{0.04}$)$_2$}
\author{M. A. Manekar, S. Chaudhary, 
M. K. Chattopadhyay, K. J. Singh, S. B. Roy$^+$ and P. Chaddah}
\address{Low Temperature Physics Laboratory,
Centre for Advanced Technology,\\ Indore 452013, India}
\date{\today}
\maketitle
\begin{abstract}
Taking the pseudobinary C15 Laves phase compound 
Ce(Fe$_{0.96}$Al$_{0.04}$)$_2$ as a paradigm for 
studying a ferromagnetic to antiferromagnetic phase transition, we present 
interesting thermomagnetic history effects in magnetotransport as well as 
magnetisation measurements across this phase transition. A comparison is 
made with history effects observed across the ferromagnetic to 
antiferromagnetic transition in R$_{0.5}$Sr$_{0.5}$MnO$_3$ crystals.
\end{abstract}
The C15 Laves phase compound CeFe$_2$, with its relatively low Curie temperature
(T$_C$  $\approx$ 230K) and reduced magnetic moment ($\approx 2.3\mu_B/$f.u.) 
\cite{1} is on the verge 
of a magnetic instability \cite{2}. Neutron measurements have shown the presence 
of antiferromagnetic (AFM) fluctuations in the ferromagnetic (FM) state of 
CeFe$_2$ below 100K \cite{3}. With small but suitable change in electronic structure
caused by doping with certain elements like Co, Al, Ru, Ir, Os, and Re at the Fe 
site \cite{4}, these AFM fluctuations get stabilized into a low 
temperature AFM state \cite{5,6,7,8}. 

While most recent experimental efforts are focussed on understanding the 
cause of this magnetic instability in CeFe$_2$ \cite{9,10}, we have recently 
addressed the question of the exact nature of this FM-AFM transition in 
Ru and Ir-doped CeFe$_2$ alloys \cite{11}. Our results show that this is a first 
order transition. The nature of the FM to AFM transition in the 
perovskite-type manganese oxide compounds R$_{0.5}$Sr$_{0.5}$MnO$_3$ 
 (R=Nd, Pr, Nd$_{0.25}$Sm$_{0.75}$) has also been the subject of close scrutiny in 
recent years \cite{12,13,14}, and has also been shown to be a first order transition.
 The existence of metastable states, which are thought to be generic to a 
first order phase transition, has been highlighted.

In this paper we report interesting thermomagnetic history dependence in 
the magnetisation and magnetoresistance in a Ce(Fe$_{0.96}$Al$_{0.04}$)$_2$ 
alloy, and 
argue that these are broader manifestations of the behaviour reported earlier 
in the perovskite-type manganese compounds \cite{12,13,14}. While the metastabilities 
can be partly explained by the phenomena of supercooling and superheating, 
we present clear signatures that the kinetics of this magnetic phase 
transition is hindered at low temperatures.

The details of the preparation and characterization of the sample can be 
found in Ref.6. The samples from the same batch have been used earlier
 in bulk magnetic, transport (Ref.6), and neutron (Ref.8) measurements. We have used 
a SQUID-magnetometer (Quantum Design MPMS5) 
for measuring magnetisation (M) as 
a function of temperature (T) and magnetic field (H).We have checked our 
results using scan length varying from 2 to 4 cm and no qualitative 
difference was found. We have used a commercial superconducting magnet and 
cryostat system (Oxford Instruments, UK) for magnetotransport measurements 
as a function of T and H. The resistivity was measured using standard dc 
four-probe technique.

The present Ce(Fe$_{0.96}$Al$_{0.04}$)$_2$  sample undergoes a paramagnetic (PM) to FM transition at
 around 200K, followed by a FM to AFM transition at around 95K (Ref.6). We 
first plot in fig. 1 M-H data at some representative T. 
The behaviour of M at T=100K
 is consistent with that of a soft (coercive field $\approx$100 Oe) FM, reaching 
technical saturation by H$\approx$3kOe. With lowering of T the nature of the M-H 
curve changes drastically with the appearance of a hysteresis bubble. 
Such hysteresis\cite{15,16}, along with the observed cubic to rhombohedral transition (Ref.8), 
have been considered earlier as possible signatures of a field-induced 
first-order metamagnetic transition from AFM to FM in Co-doped CeFe$_2$ (Ref.15). 
At T=5K, we find that if the maximum field excursion is less than $|30|$ kOe, 
the M-H curve remains reversible in the high field regime. 
In this field regime the sample remains in the AFM state.(The observed nonlinearity in $\pm$5 kOe 
regime is due to parasitic weak ferromagnetism \cite{17} leading to a canted 
spin state \cite{8,16,18}) When the applied H is increased beyond H$_M\approx $30 kOe, 
M rises rapidly and upon lowering H a hysteresis is observed. The hysteresis loop,
however, collapses before H is reduced to zero (coercive field $\approx$ 300 Oe),
 and reappears again in the third quadrant, giving rise to a butterfly-like
 hysteresis loop. In fig. 2 we 
present resistivity ($\rho$) as a function of H, 
at the representative temperatures T= 3K, 5K, and 20K. The sample is 
initially cooled to each temperature in zero field. 
We see the clear signature of a field induced AFM-FM transition at a field H$_M$(T), 
where the resistivity decreases sharply with increasing H. 
It is to be noted that in the present  case of Ce(Fe$_{0.96}$Al$_{0.04}$)$_2$ 
both FM and AFM phases are metallic,  and the 
change in resistivity observed is not as drastic as 
in R$_{0.5}$Sr$_{0.5}$MnO$_3$ with R=Nd \cite{12}. On reducing the field from 
well above H$_M$, a clear hysteresis is seen in the $\rho$ vs H plot (see figures 
2a to 2c). This hysteresis is due to metastable states expected across a 
first order transition where the inequality between the free energies of the two phases 
changes sign on a (H$_M,T_N$) line, but the transition to the higher 
entropy phase actually occurs on a (H$^{**},T^{**}$) line because of superheating, 
and the transition to the lower entropy phase actually 
occurs on a (H$^*,T^*$) 
line because of supercooling\cite{19}. We accordingly attribute this hysteresis 
in M vs H and in $\rho$ vs H to the first order nature of the field induced 
FM-AFM transition. Similar hysteresis observed in the 
$\rho$-H plots of R$_{0.5}$Sr$_{0.5}$MnO$_3$ 
has also been attributed to the first order nature 
of the phase transition \cite{12,13,14}. 

In this picture, the FM state continues to exist as a supercooled metastable 
state when H is lowered isothermally below H$_M$, upto the limit H$^*$. Between 
H$_M$ and H$^*$ fluctuations can help in nucleating droplets of the AFM state, 
and at H$^*$ an infinitesimal fluctuation will drive the whole system to the 
stable AFM phase. Heterogeneous nucleation can thus cause a spatial 
distribution of the field till which supercooling is actually observed 
at T, resulting in the (H$^*,T^*$) line getting broadened into a band.
 Early theoretical arguments\cite{20} and recent measurements\cite{21} 
also showed that a sample with disorder can have a spatial distribution of the phase 
transition field H$_M$ in a general first order transition, 
hence broadening the (H$_M$,T$_N$) line into a band. 
This disorder would also cause the (H$^*,T^*$) 
and (H$^{**},T^{**}$) lines to be broadened into bands. This is depicted 
schematically in fig. 3a, and is consistent with earlier neutron 
scattering observation in Al-doped CeFe$_2$ of coexisting FM and AFM phases 
over a finite temperature regime \cite{8} .

We now come to some interesting features seen at very low temperatures, 
where the effect of thermal fluctuations is reduced. As seen in figs. 2b
and 2c, when the applied field is reduced to zero from H$_{max}$ well above H$_M$
at T=3 and 5K, the 
$\rho$(H=0) lies distinctly below the initial ZFC $\rho$(H=0), thus giving rise to 
an open hysteresis loop. This kind of open hysteresis loop has been 
reported earlier for single crystal Nd$_{0.5}$Sr$_{0.5}$MnO$_3$ 
samples at low temperature \cite{12}.
We attribute this lower resistance to the existence of a residual 
metastable FM phase even at H=0. Is (H=0,T=3K) within the (H$^*$,T$^*$) band,
or can such a residual FM phase persist below the (H$^*$,T$^*$) band? We shall
return to this question.

The envelope $\rho$-H hysteresis curves shown in fig. 2 are obtained by reducing 
the field from H$_{max}$ to zero to -H$_{max}$, 
and raising it back to H$_{max}$.  
We see in fig 2 that the virgin $\rho$-H curve lies outside the envelope 
hysteresis curve. As seen in fig 1,  the virgin M-H curve 
at 5K also lies clearly outside 
the envelope hysteresis curve. We argue from existing data that this 
anomalous feature would also be seen in 
Nd$_{0.5}$Sr$_{0.5}$MnO$_3$ at T$<$20K. The reported 
$\rho$-H curve (see fig.2B to 2D of Ref. 12) exhibits an open hysteresis loop. 
It appears obvious that if H were again increased fron zero to 120 kOe, 
the forward leg of this envelope curve would merge with the virgin 
$\rho$-H curve only at some finite (and large) field. We consider this 
anomalous relation between virgin and envelope hysteresis curves at 
low temperatures, seen in both magnetotransport and magnetisation 
measurements, as significant.

We now present in fig.4, $\rho$ vs T plots in fields of H=0, 5, 20, and 30 kOe.
In each case we have cooled the sample to 5K in zero field  and then 
applied H at this temperature. Resistivity is then measured as the sample is warmed 
well into the FM state. The sample is then cooled back to 5K in the field H,
 allowing a measurement of thermal hysteresis. The appearance of  magnetic 
superzones\cite{6,7} at the AFM-FM transition gives rise to the distinct structure 
observed in  $\rho$-T. There is a marked hysteresis between the warming and 
cooling cycles, because the FM (or AFM) 
phase can be supercooled (or superheated) and exists as a metastable phase 
between the (H$_M$,T$_N$) line and the (H$^*,T^*$) line 
(or the (H$^{**},T^{**}$) line). 
For H=0 and 5 kOe, the FM to AFM transition is completed during cooldown
 at T$\approx$20K; this indicates that (H=0,T=5K) 
and (H=5kOe, T=5K) points lie below the (H$^*$,T$^*$) band and 
no supercooled FM phase is expected to be metastable at H=0 at T=3K or 5K. 
This is in striking contrast with the data in fig 2 which shows that the
 resistance of the AFM state is not restored when H is reduced to zero 
isothermally.(The  $\rho$ vs T data in R$_{0.5}$Sr$_{0.5}$MnO$_3$ shows the same contrast with 
the  $\rho$-H data \cite{12,14}) The data in fig. 4 also shows 
that for H=20 kOe and H=30 kOe 
the resistivity does not rise to its full AF state value on the cooling curve down
 to 5K, even though the FM to AFM transformation appears to have been
arrested at around 15K. (This is again similar to the observations in Ref. 14 on 
the single crystal manganite samples.)

We now summarize the unusual findings of the present study.
\begin{enumerate}
\item The envelope $\rho$-H curve at 5K and 3K (fig.2) does not return at H=0 to the 
virgin curve value of $\rho$(H=0), while the FM to AFM transition is complete 
when the sample is cooled to these temperatures in low field. A similar 
behaviour is seen from the single crystal studies 
on R$_{0.5}$Sr$_{0.5}$MnO$_3$ in another first order 
FM to AFM transition\cite{12,14}.
\item The butterfly $\rho$-H and M-H hysteresis loops have an anomalous virgin curve
 at low temperatures, in that the virgin curve lies outside the envelope 
hysteresis curve in both measurements.
\item In the field-cooled measurement of $\rho$ vs T at H=20 kOe, and 30 kOe, 
the FM to AFM transition appears to be arrested at about 15K even while the 
transformation is incomplete, and remains incomplete down to 5K. 
These results are supported by the study of T dependence of M in both the
zero-field-cooled and field-cooled mode \cite{22}.
Similar behaviour in the the resistivity studies are seen in 
the single crystal studies of R$_{0.5}$Sr$_{0.5}$MnO$_3$ \cite{14}. 
\end{enumerate}

As a possible explanation we introduce the idea that the kinetics of the FM 
to AFM transition gets hindered at low T and in high H. 
We concentrate on the (H$^*$,T$^*$) band 
in fig. 3b, and recognize that for (H,T) values below this band the free 
energy barrier separating the FM from the AFM phase has dropped to zero 
throughout the sample \cite{19}. An infinitesimal fluctuation should drive any 
FM region to the AFM phase. But all our observations indicate that at very 
low T the unstable FM regions remain in the AFM phase.
It is well known that at sufficiently low T the characteristic
time for structural relaxation becomes larger 
than experimental time-scales \cite{23}. 
We postulate that at sufficiently low T
 the displacive motion of atoms involved in the structural distortion 
that is associated with the FM-AFM 
transition in  Ce(Fe$_{0.96}$Al$_{0.04}$)$_2$ (Ref. 8), 
becomes negligible on experimental time scales. The high temperature-high 
field FM phase is then frozen-in.
We accordingly postulate that below a certain temperature 
T$_K$(H) the kinetics of the FM-AFM transformation is hindered and arrested just like 
in a quenched metglass. (This is similar to observations at high pressures
where the high density phase cannot transform\cite{24} to the low density 
phase below a certain T$_K$, with T$_K$ rising as the pressure rises.) 
We would depict this as a (H$_K$,T$_K$) line in the 
two-control-variable (H,T) 
space, which we broaden into a band with the same argument used to replace 
the other thermodynamic transition lines by a band.   
At temperatures below the band the freezing-in occurs throughout the sample; 
within the band it occurs in some regions of the sample. With this
 conjecture we now explain the three unusual findings enumerated above. 
We use the schematic in fig 3b, where paths labeled by 1 and 2 indicate
 cooling in constant field and lowering H at constant T, 
respectively. We assume that we start always with a sample that is prepared 
to be completely in the FM phase by warming (or increasing field) to 
a point well above the (H$^{**},T^{**}$) band.

Let us cool sequentially to points A, B, F, and G along path 1. At point A we 
observe FM and AFM coexistence (with FM being metastable) while at point B 
the entire sample is in the AFM phase. The (H$_K$,T$_K$) band has no observable
effect as we cool to points F or G. This corresponds to our $\rho$-T data at 
H=0 or 5 kOe. Following path 1 again, we cool in higher fields to reach, 
sequentially, points C, D, E, and L. At C we have two phase coexistence 
with FM transforming to AFM as temperature is lowered. This transformation 
is arrested at D, and the FM fraction is frozen-in at E and L, even though 
it would have kept reducing in an ergodic system. Thus we have a frozen-in 
FM phase at L even though it is unstable. This explains the field-cooled $\rho$ 
vs T and M vs T (Ref.22) data at fields of 20 and 30 kOe. We now follow path 2 and 
lower the field sequentially to points C and B. At point C we see two-phase 
coexistence and at point B the sample is fully AFM. This explains our M-H 
data and our $\rho$-H data at higher T. We now follow path 2 and lower 
the field to points E and F. At point E the FM phase is frozen-in 
throughout, while at F some regions of the sample are no longer kinetically 
arrested and transform to the AFM phase. This corresponds to our $\rho$-H data at
 3K and 5K, with only part recovery of the AFM phase even though path 1 
gives a full recovery of the  AFM  phase. And in the ZFC state at H=0 we 
reached point F by path 1. This also explains the anomalous virgin curves 
because the virgin curve starts at F after path 1, then goes above the 
(H$^{**}$,T$^{**}$) band and returns to point F by path 2 thereby retaining a fraction 
of FM phase. The forward hysteresis curve now starts with coexisting 
FM-AFM phases unlike the virgin curve which had only AFM phase. 
Because of the larger M and lower $\rho$ of the FM phase, the forward hysteresis 
envelope thus lies above (or below) the virgin curve in isothermal 
M-H (or $\rho$-H) measurements. Finally, if we follow path 2  and reduce 
the field isothermally to reach point G, then the FM phase is 
frozen-in completely, and no AFM phase is recovered. This explains the 
$\rho$-H data at 10K reported (Ref.14) in the R$_{0.5}$Sr$_{0.5}$MnO$_3$
single crystal with R=Nd$_{0.25}$Sm$_{0.75}$.

To conclude, we have observed unusual history effects in magnetization and 
magnetotransport measurements across the FM-AFM transition 
in Ce(Fe$_{0.96}$Al$_{0.04}$)$_2$, and 
have discussed similarities with earlier single crystal data on
R$_{0.5}$Sr$_{0.5}$MnO$_3$ across another 
first order FM-AFM transition. We have argued that the kinetics of this 
FM-AFM transition is hindered at low T. This observation may
 be of  relevance to other first order transitions where it is more
 difficult to vary two control variables; one example is the 
high density amorphous water (HDAW)  to low density amorphous water (LDAW) 
transition which is observed with reducing pressure at 130K, but whose
 kinetics appears to be arrested at 77K \cite{23,25}. 
These FM-AFM transitions can be 
used as paradigms to study various interesting aspects of a first order 
transition like nucleation and growth, supercooling and superheating, 
hindered kinetics, etc in a relatively easy and reproducible manner. 

+ Corresponding author

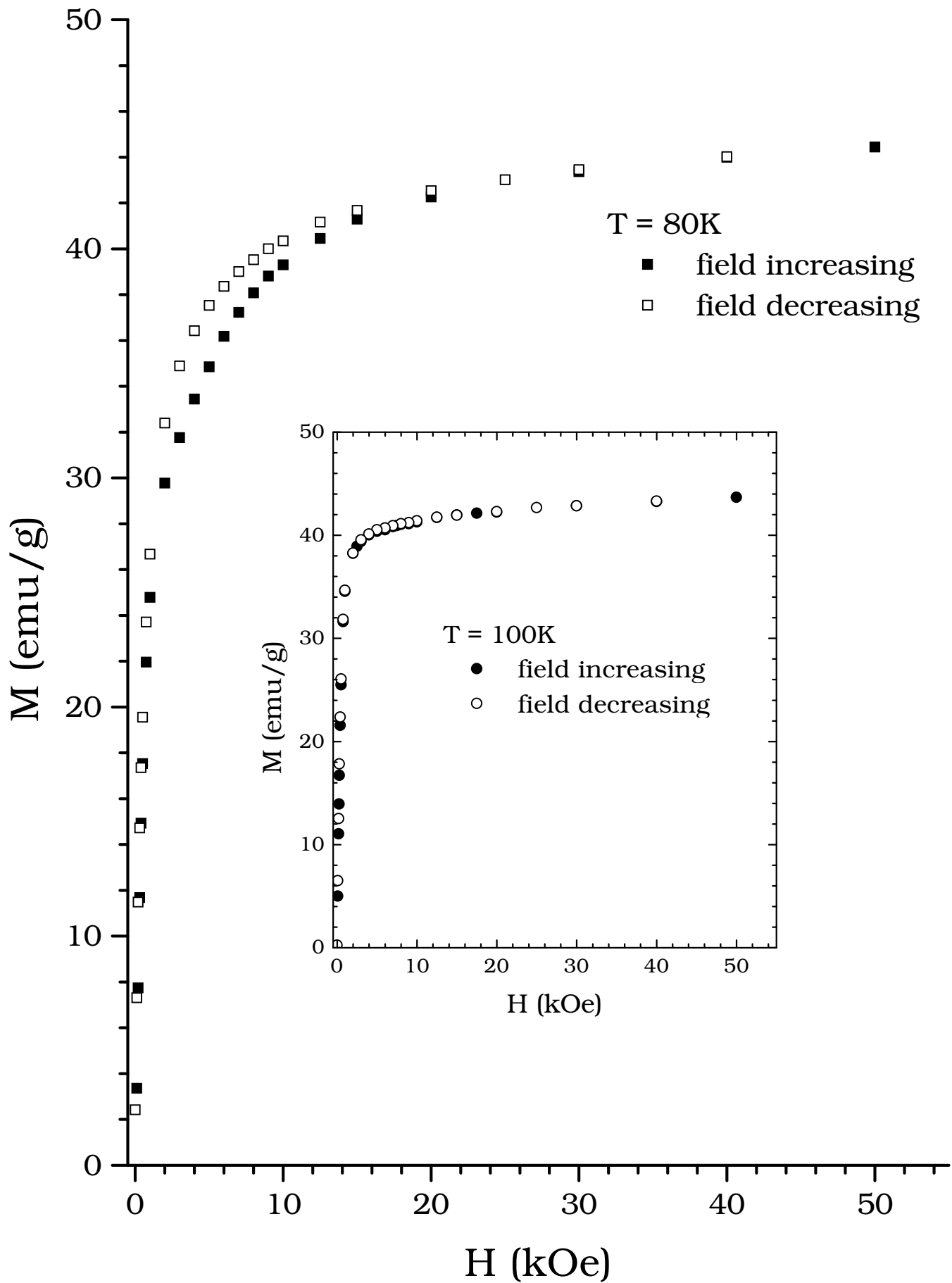
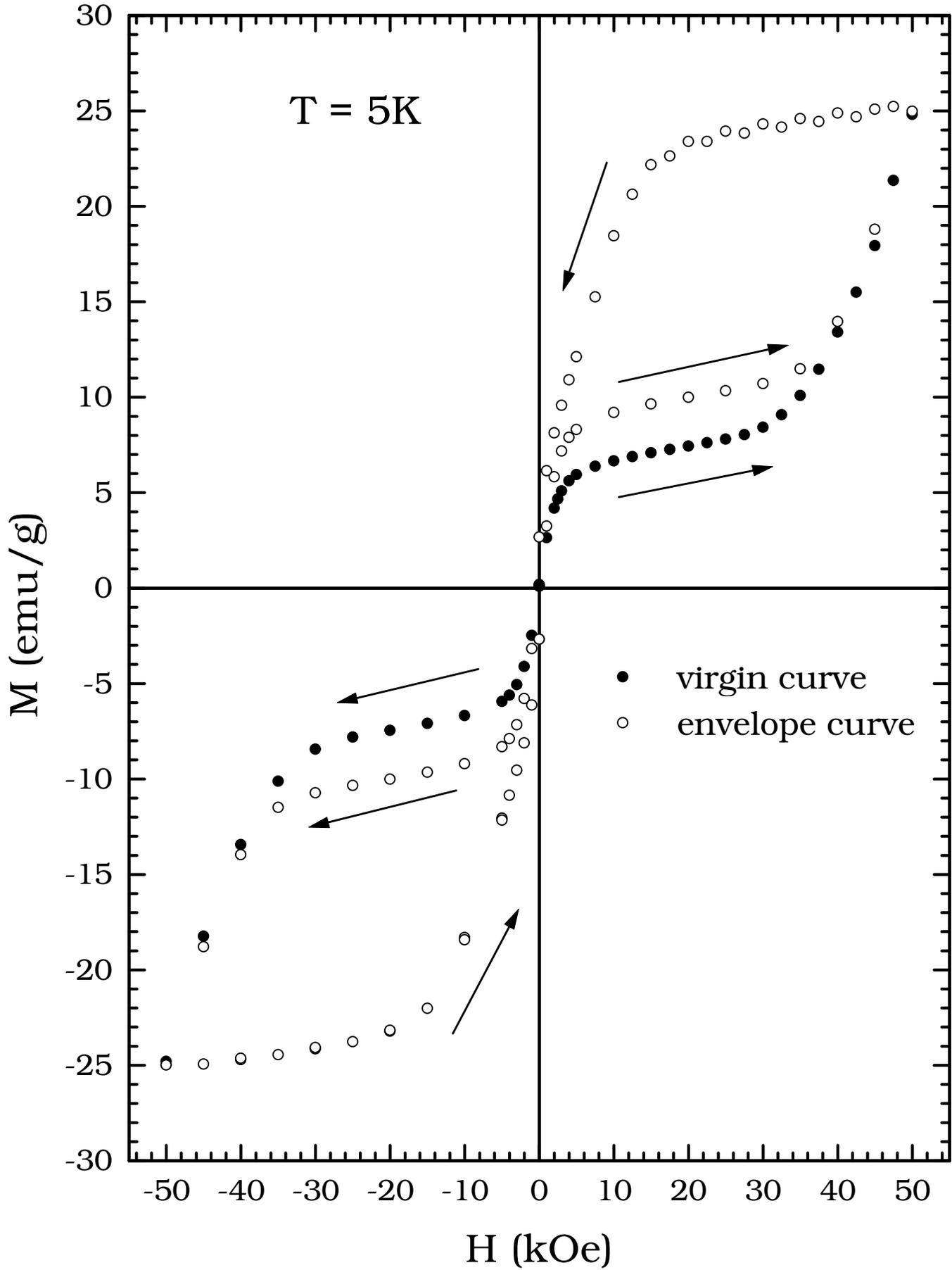
\begin{figure}
\caption{M vs H plots of Ce(Fe$_{0.96}$Al$_{0.04}$)$_2$ obtained after 
cooling in zero field 
(a) at T=80K and 120K (b) at T=5K. Note that
at T=5K the virgin M-H curve lies outside
the envelope M-H curve. To confirm this anomalous nature of virgin curve
we have also drawn this in the negative field direction after zero field
cooling the sample.}
\end{figure} 
\begin{figure}
\caption{R vs H plots of Ce(Fe$_{0.96}$Al$_{0.04}$)$_2$ 
at T=20K, 5K and 3K. Dashed lines represent virgin curve
drawn in the negative field direction after zero field
cooling the sample. }
\end{figure} 
\begin{figure}
\caption{(a)Schematic representation of broadened band 
of phase transition (H$_M$,T$_N$),
supercooling (H$^*$,T$^*$) and superheating (H$^{**}$, T$^{**}$) lines. The
last two present the limits of metastability. See text for details.
(b)Schematic representation of the relative position of the band (H$_K$,T$_K$)
(across which the kinetics of 
FM to AFM transformation is hindered) with respect to 
(H$^*$,T$^*$). See text for details.}
\end{figure}
\begin{figure}
\caption{$\rho$ vs T of Ce(Fe$_{0.96}$Al$_{0.04}$)$_2$ 
plots showing the FM-AFM transition with H=0, 5 kOe, 20 kOe and 30 kOe. 
Inset shows the zero field $\rho$-T data with both PM-FM and FM-AFM transition.
}
\end{figure}
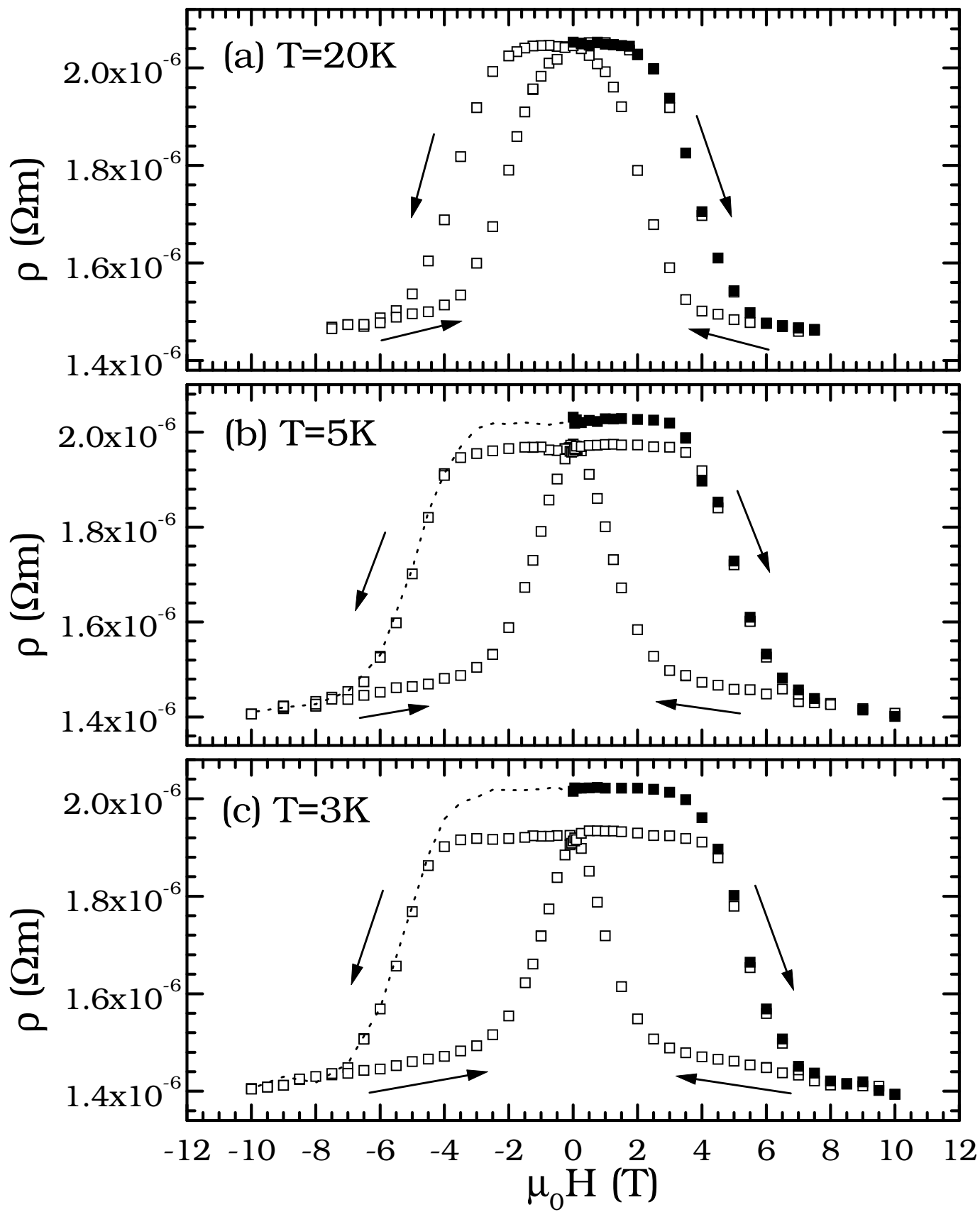
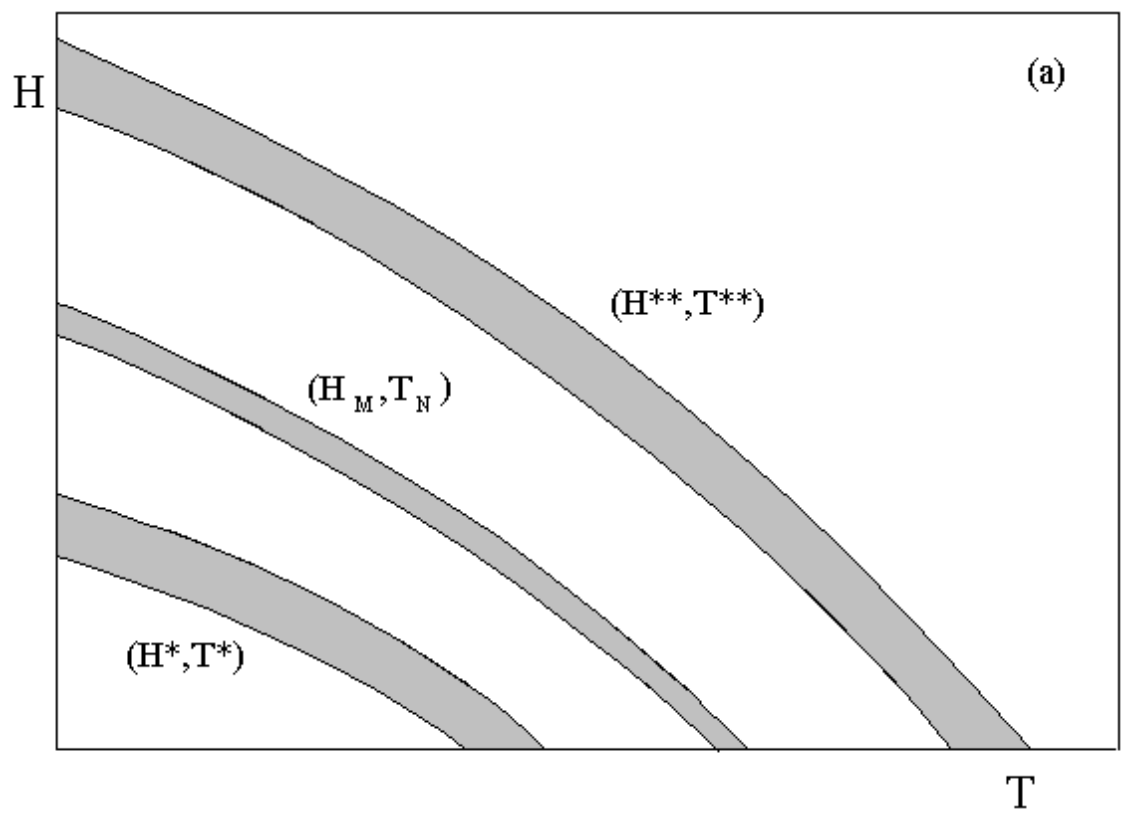
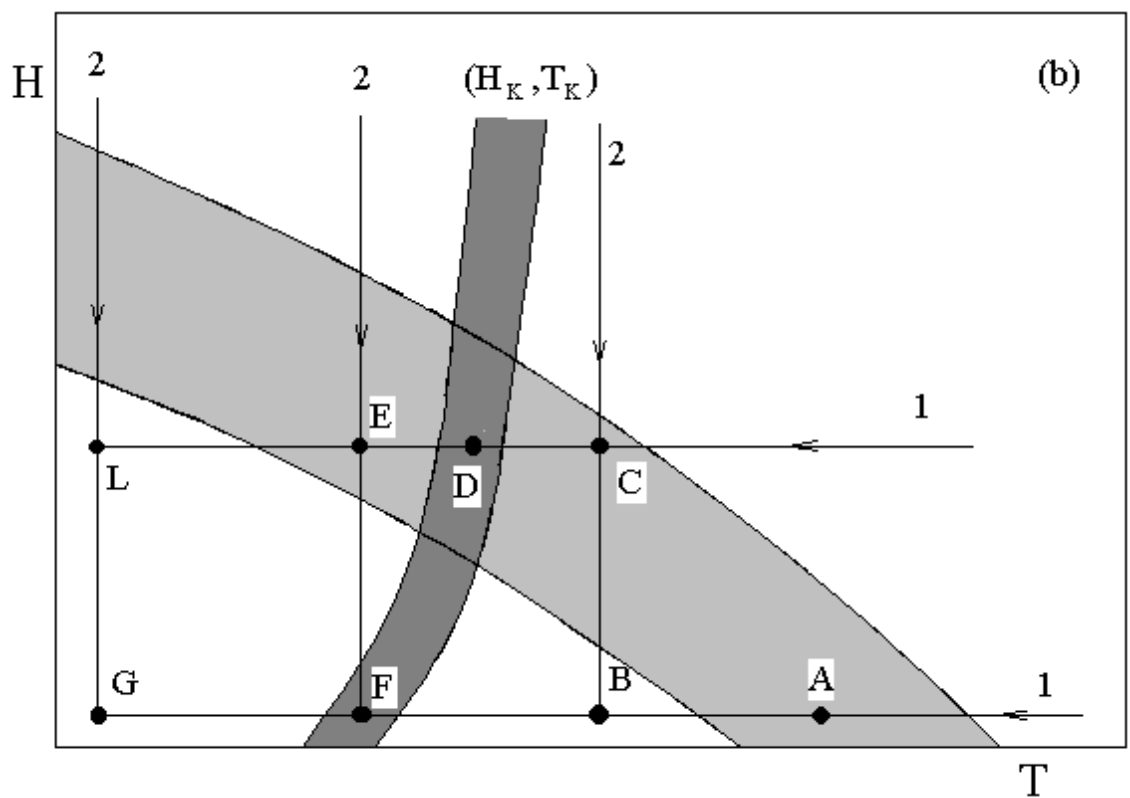
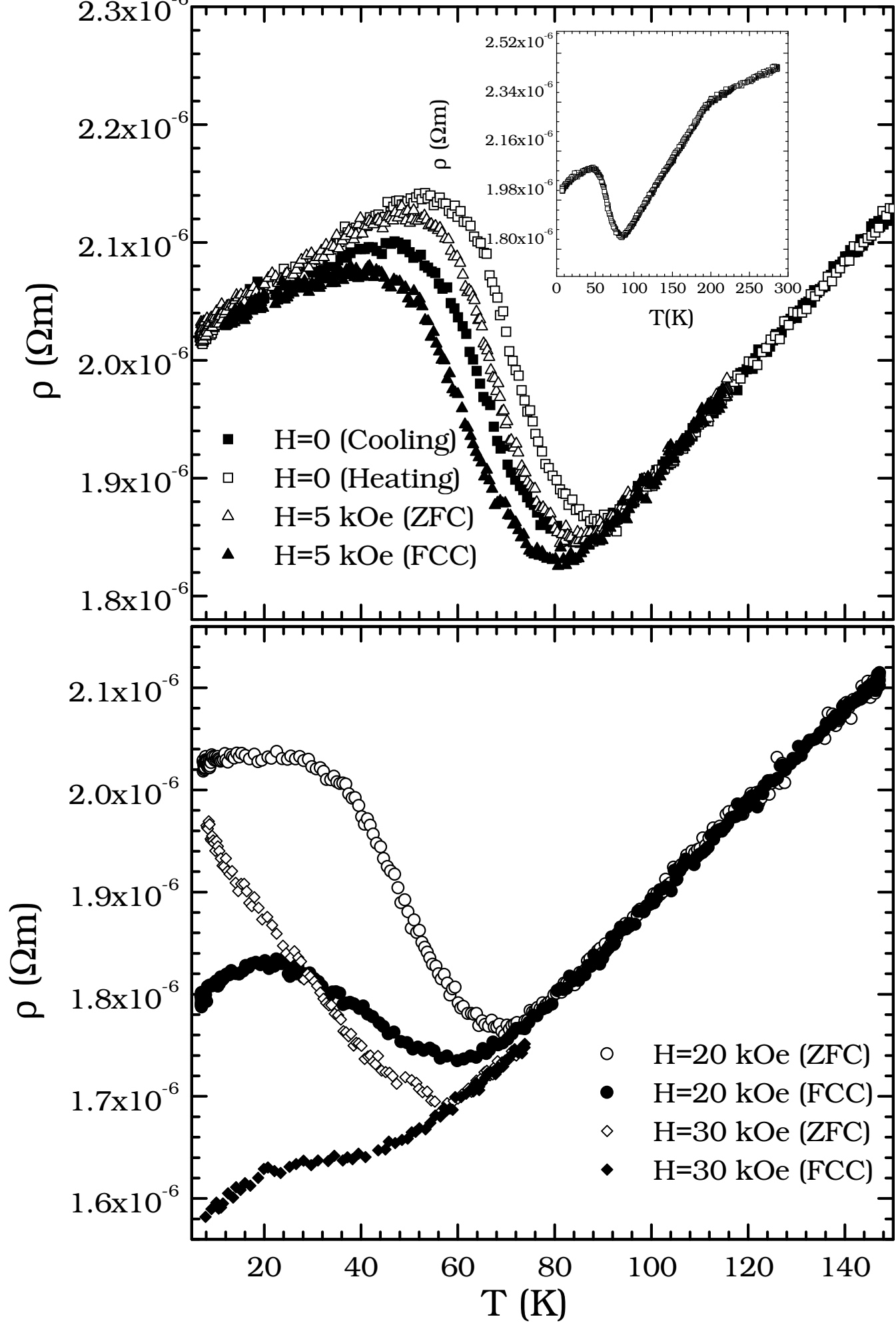
\end{document}